\documentclass[]{spie}  

\usepackage[perpage]{footmisc}
\usepackage{subcaption} 
\usepackage{amsmath,amsfonts,amssymb}
\usepackage{graphicx}
\usepackage[colorlinks=true, allcolors=blue]{hyperref}
\usepackage[numbers]{natbib}

\title{Single electron Sensitive Readout (SiSeRO) X-ray detectors: Technological progress and characterization}

\author[a]{Tanmoy Chattopadhyay}
\author[a]{Sven Herrmann}
\author[a]{Peter Orel}
\author[a,b]{R. G. Morris}
\author[a]{Daniel R. Wilkins}
\author[a,b,c]{Steven W. Allen}
\author[d]{Gregory Prigozhin}
\author[d]{Beverly LaMarr}
\author[d]{Andrew Malonis}
\author[d]{Richard Foster}
\author[d]{Marshall W. Bautz}
\author[e]{Kevan Donlon}
\author[e]{Michael Cooper}
\author[e]{Christopher Leitz}

\affil[a]{Kavli Institute of Particle Astrophysics and Cosmology, Stanford University, 452 Lomita Mall, Stanford, CA 94305, USA}
\affil[b]{SLAC National Accelerator Laboratory, 2575 Sand Hill Road, Menlo Park, CA 94025, USA}
\affil[c]{Department of Physics, Stanford University, 382 Via Pueblo Mall, Stanford CA 94305, USA}
\affil[d]{Kavli Institute for Astrophysics and Space Research, Massachusetts Institute of Technology, Cambridge, MA USA}
\affil[e]{MIT Lincoln Laboratory, Lexington, MA, USA}

\authorinfo{Send correspondence to T. Chattopadhyay (tanmoyc@stanford.edu)}

\begin{document} 
\maketitle

\begin{abstract}
Single electron Sensitive Read Out (SiSeRO) is a novel on-chip charge detector output stage for charge-coupled device (CCD) image sensors. Developed at MIT Lincoln Laboratory, this technology uses a p-MOSFET transistor with a depleted internal gate beneath the transistor channel. The transistor source-drain current is modulated by the transfer of charge into the internal gate. At Stanford, we have developed a readout module based on the drain current of the on-chip transistor to characterize the device. Characterization was performed for a number of prototype sensors with different device architectures, e.g. location of the internal gate, MOSFET polysilicon gate structure, and location of the trough in the internal gate with respect to the source and drain of the MOSFET (the trough is introduced to confine the charge in the internal gate). Using a buried-channel SiSeRO, we have achieved a charge/current conversion gain of $>$700 pA per electron, an equivalent noise charge (ENC) of around 6 electrons root mean square (RMS), and a full width half maximum (FWHM) of approximately 140 eV at 5.9 keV at a readout speed of 625 Kpixel/s. In this paper, we discuss the SiSeRO working principle, the readout module developed at Stanford, and the characterization test results of the SiSeRO prototypes. We also discuss the potential to implement Repetitive Non-Destructive Readout (RNDR) with these devices and the preliminary results which can in principle yield sub-electron ENC performance. Additional measurements and detailed device simulations will be  essential to  mature the SiSeRO technology. However, this new device class presents an exciting technology for next generation astronomical X-ray telescopes requiring fast, low-noise, radiation hard megapixel imagers with moderate spectroscopic resolution.
\end{abstract}



\section{INTRODUCTION}
\label{sec:intro}  
X-ray Charge Coupled Devices (CCDs) \citep{Lesser15_ccd,gruner02_ccd} 
have been the primary detector technology for soft X-ray instrumentation for more than three decades in X-ray astronomy. The low noise performance ($<$4 $e^{-}_{RMS}$), broad energy response up to tens of keV, high spatial and moderate energy resolution made them an obvious choice for the X-ray spectro-imagers in missions such as ASCA, Chandra, XMM-Newton, Swift, Integral, Hitomi, and AstroSat. Chandra, in particular, has been extremely successful in revealing the detailed properties of X-ray astronomical sources by combining the high angular resolution imaging with the excellent detection properties of CCDs. 

Next generation X-rays CCDs, with higher readout speed and improved noise performance, also hold the potential to support future astronomical missions such as the Lynx Observatory \citep{gaskin15_lynx}, proposed earlier to NASA's 2020 Decadal Survey, and AXIS probe class mission\footnote{https://axis.astro.umd.edu/}, which will have order-of-magnitude larger collecting areas to probe deeper into the high redshift, faint X-ray universe. Noise performance as low as 1 $e^{-}_{RMS}$ and associated improvements in low energy response ($<$500 eV) will also be highly beneficial to fully utilize these observatories. MIT Lincoln Laboratory (MIT-LL), MIT and Stanford University (SU) have together made substantial improvements in X-ray CCD technology in recent years \citep{bautz18,bautz19,bautz20}, and in the development of fast, low noise readout electronics to support these detectors \citep{chattopadhyay22_ccd,herrmann20_mcrc,Bautzetal2022,Oreletal2022}. In parallel,
MIT-LL has been developing a unique Single electron Sensitive readout stage (hereinafter SiSeRO) for CCDs, intended to provide even greater responsivity and significantly better noise performance even at readout speeds $>$2 MHz. 

In Chattopadhyay et al. 2022 \citep{chattopadhyay22_sisero}, we discussed the working principle of SiSeROs and demonstrated the first results from these devices using a readout module developed at SU. The working principle of SiSeROs is similar in some respects to DEPFET detectors \citep{kemmer87_depfet,strueder00_depfet_imager} and draws on earlier work on floating-gate amplifiers described in Matsunaga et al. 1991 \citep{matsunaga91}. SiSeRO devices employ a p-channel metal-oxide-semiconductor field-effect transistor (MOSFET) straddling the n-channel of the CCD. When a charge packet is transferred to the internal channel beneath the p-MOSFET, it modulates the transistor drain current, which can be sensed directly. For our first generation prototypes, we obtained an equivalent noise charge (ENC) of 15 $e^{-}_{RMS}$. In this paper, we 
update the characterization test results for a new SiSeRO device (CCID93) with a buried-channel p-MOSFET transistor, obtaining highly encouraging results. We have also developed techniques to suppress 1/f correlated noise using an optimized digital filter when measuring the charge signal in the pixels. In addition, we have for the first time demonstrated Repetitive Non-Destructive Readout (RNDR) in these detectors, wherein the charge signal is measured multiple times by moving the charge back and forth non-destructively. In principle this can reduce the noise well below the 1/f barrier and into the sub-electron regime. In future work, we plan to continue to mature the SiSeRO design and further optimize the digital filtering and RNDR.    

In Sec. \ref{sec:sisero}, we give a brief introduction to the SiSeRO amplifiers. This is followed in Sec. \ref{sec:results} by a short description of the characterization test stand, readout module and the new test results for the CCID93 buried-channel SiSeRO. In Sec. \ref{sec:digfiltering} and Sec. \ref{sec:rndr}, we discuss the new analysis techniques, namely the digital filtering and RNDR techniques respectively. The results are summarized along with future plans in Sec. \ref{summary}.


\section{Overview of SiSeRO devices}\label{sec:sisero}

\begin{figure}
    \centering
    \includegraphics[width=0.75\linewidth]{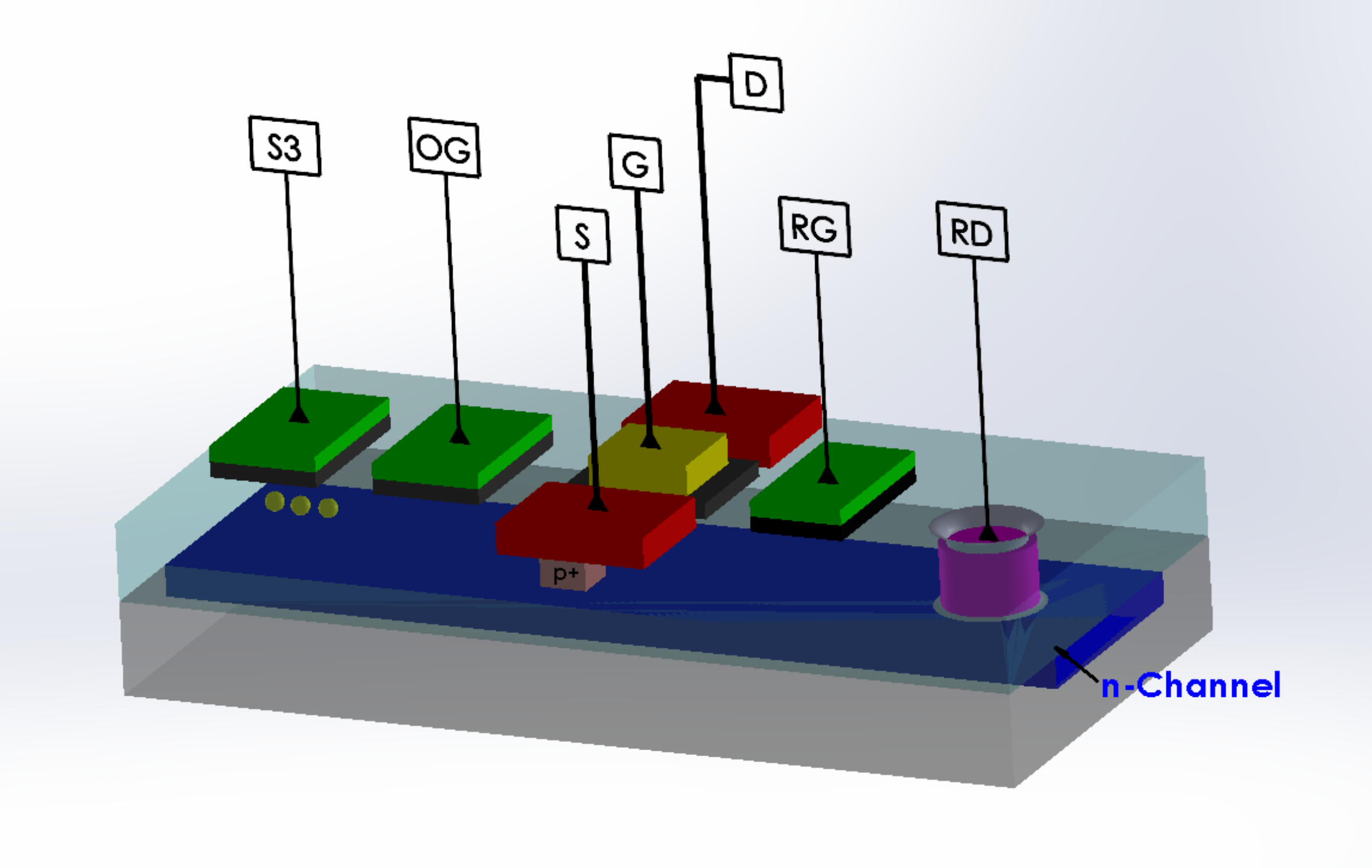}
    \caption{Working principle and schematic of the SiSeRO. It uses a p-MOSFET transistor with an internal gate beneath the p-MOSFET. Charge transferred from the last serial gate (S$_3$) of the serial register to the internal gate through the output gate (OG) modulates the drain current of the p-MOSFET. See text for more details. Picture courtesy \cite{chattopadhyay22_sisero}.}
    \label{sisero}
\end{figure}
In this section, we give a brief overview of the SiSeRO devices we are testing. Details of the working principle can be found in \citep{chattopadhyay22_sisero}. The SiSeRO amplifier, shown in Fig. \ref{sisero}, is a floating-gate amplifier, where a P-MOSFET gate (`S', `G', and `D' in the figure refer to Source, Gate, and Drain respectively) sits above the CCD channel. When a CCD charge packet is transferred beneath the gate, it modulates the drain current of the transistor, which is proportional to the source signal strength. The signal is brought directly out of the package through the source or drain of the transistor, enabling the device to be operated in a drain-current readout mode. Once the charge is readout, the internal channel is emptied by the resetting the channel to reset drain (`RD') using reset gate (`RG') switch. The `S$_3$' and `OG', in the figure refer to the last serial clock register and output gate respectively. The output gate is normally connected to a small positive DC bias to shift the charge from the serial register to the internal channel.   

For current readout devices like SiSeROs, the noise of the output stage depends on the conversion gain, $G_q$ (pA/electron) and the transconductance, $g_m$ ($\mu$S) of the transistor. From first principles, it can be estimated that the device output stage noise can be as low as 1 $e^{-}_{RMS}$ for a $G_q$ of 1400 pA/e and $g_m$ of around 20 $\mu$S for 1 MHz of readout speed. In the case of SiSeROs based on a single polysilicon MOSFET, the internal gate minimizes parasitic capacitance on the sense node, resulting in 
a high conversion gain and minimized noise. 
Further, MOSFETs show larger transconductance per unit area than JFETs (used in CCDs) and especially small MOSFETs can achieve very high signal speeds. 
SiSeROs also offer the potential for RNDR: since the charge packet in the internal gate is unaffected by the read out process, it can be moved around like any charge packet in a CCD. This can in principle reduce the noise below the 1/f barrier and deep into the sub-electron regime. 
Together, these properties identify SiSeROs as a device class with the potential for sub-electron noise and very high-speed performance.

In Chattopadhyay et al. 2022 \citep{chattopadhyay22_sisero}, we used the first generation prototype SiSeRO devices to demonstrate its working principle and presented the first results using a drain current readout module. We estimated a charge conversion gain of 700 pA/e and around 20 $\mu$S of transconductance, obtaining an ENC of 15 $e^{-}_{RMS}$ and a full width at half maximum (FWHM) spectral resolution of 230 eV at 5.9 keV. These first prototypes used surface channel transistors in which we expect the Si interface states to capture and release mobile carriers, causing excess noise that is multiple times that of the theoretical lower limit. In this paper, we characterize a new SiSeRO device that uses a buried transistor channel. In absence of trapping and de-trapping of the charge carriers from the Si interface states, we can expect improved noise performance from these devices. Details of the experiment and results are given in the next sections.


\section{Characterization of a CCID93 device (buried-channel SiSeRO)}\label{sec:results}

\begin{figure}
    \centering
    \begin{subfigure}{.45\textwidth}
    \includegraphics[width=\linewidth]{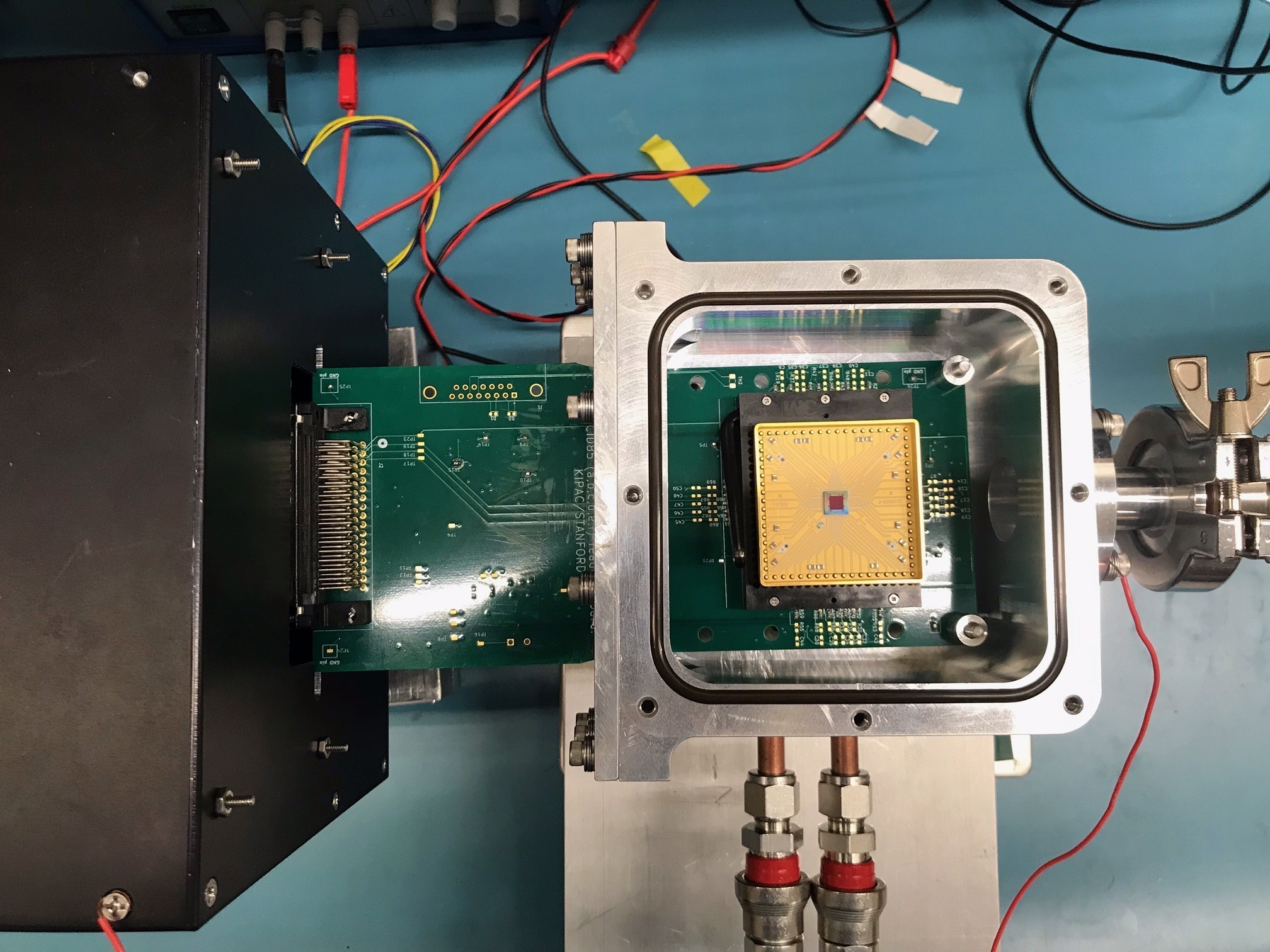}
    \caption{}
    \end{subfigure}
     \begin{subfigure}{.52\textwidth}
    \includegraphics[width=\linewidth]{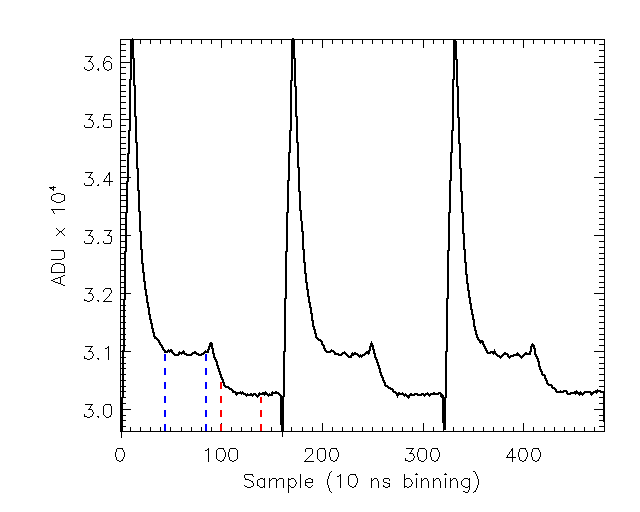}
    \caption{}
    \end{subfigure}
    \caption{(a) The experimental set up with a CCID93 bured channel SiSeRO device mounted inside the chamber. A beryllium window mounted on the top flange (not shown here) is used for X-ray entrance window. (b) SiSeRO output video waveform (shown for 3 pixels here) obtained from the Archon controller. The regions within the blue and red dashed lines represent the baseline and signal levels respectively. The difference of the two levels is proportional to the source signal.}
    \label{setup}
\end{figure}
\subsection{Experiment Test Stand}
The test stand, along with the readout module (also known as the `Tiny Box', see \citep{chattopadhyay20_spie} for details) and the X-ray detector are shown in Fig. \ref{setup}a. A compact (13 cm $\times$ 15 cm $\times$ 6.5 cm) aluminum vacuum chamber houses the X-ray detector. The detector is mounted in the middle of the chamber on an aluminum cold block (or cold finger) and faces upward towards the top flange (not shown here). An X-ray entrance window on the top flange allows X-ray photons to illuminate the detectors. A thermo-electric cooler (TEC) is used to cool down the detectors. A liquid plate on the back of the bottom flange removes the heat deposited by the TEC. We use a Proportional-Integral-Derivative (PID) algorithm to control the temperature of detectors with better than 0.1$^\circ$C accuracy. 

For this experiment, we used a CCID93 detector (see Fig. \ref{setup}a), a prototype X-ray CCD with a buried-channel SiSeRO at the output stage. The test device, developed by MIT-LL, is fabricated in an n-channel, low-voltage, single-poly process. The device active area is $\sim$4 mm $\times$ 4 mm in size with a 512 $\times$ 512 array of 8 $\mu$m pixels. The test detector, used in the experiment, is a front-illuminated device.
The readout module consists of a preamplifier board (the green circuit board in Fig. \ref{setup}a) and an Archon controller (the black box in the figure). The preamplifier uses a drain current readout where an I2V amplifier first converts and amplifies the SiSeRO output and a differential driver, at the second stage, converts the output to fully differential signal at the input of the differential ADCs in the Archon controller. In Chattopadhyay et al. 2022 \citep{chattopadhyay22_sisero}, we discuss the schematic of the preamplifier chain along with LTspice simulations\footnote{https://www.analog.com/en/design-center/design-tools-and-calculators/ltspice-simulator.html} for the transient and ac performance of the circuit. 
The simulation results suggest a low noise yield from the readout circuit. The Archon \citep{archon14}, procured from Semiconductor Technology Associates, Inc (STA\footnote{http://www.sta-inc.net/archon/}), provides the necessary bias and clock signals to run the detector, digitize the output signal and perform correlated double sampling (CDS) to generate 2D images. 

\subsection{Results}
\begin{figure}
    \centering
    \begin{subfigure}{.48\textwidth}
    \includegraphics[width=\linewidth]{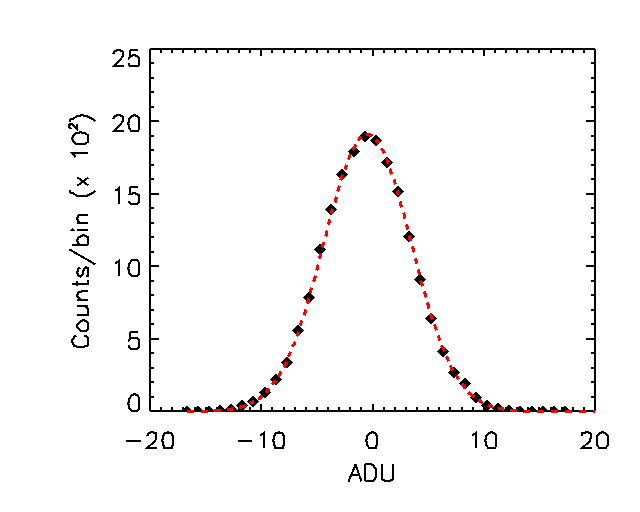}
    \caption{}
    \end{subfigure}
     \begin{subfigure}{.505\textwidth}
    \includegraphics[width=\linewidth]{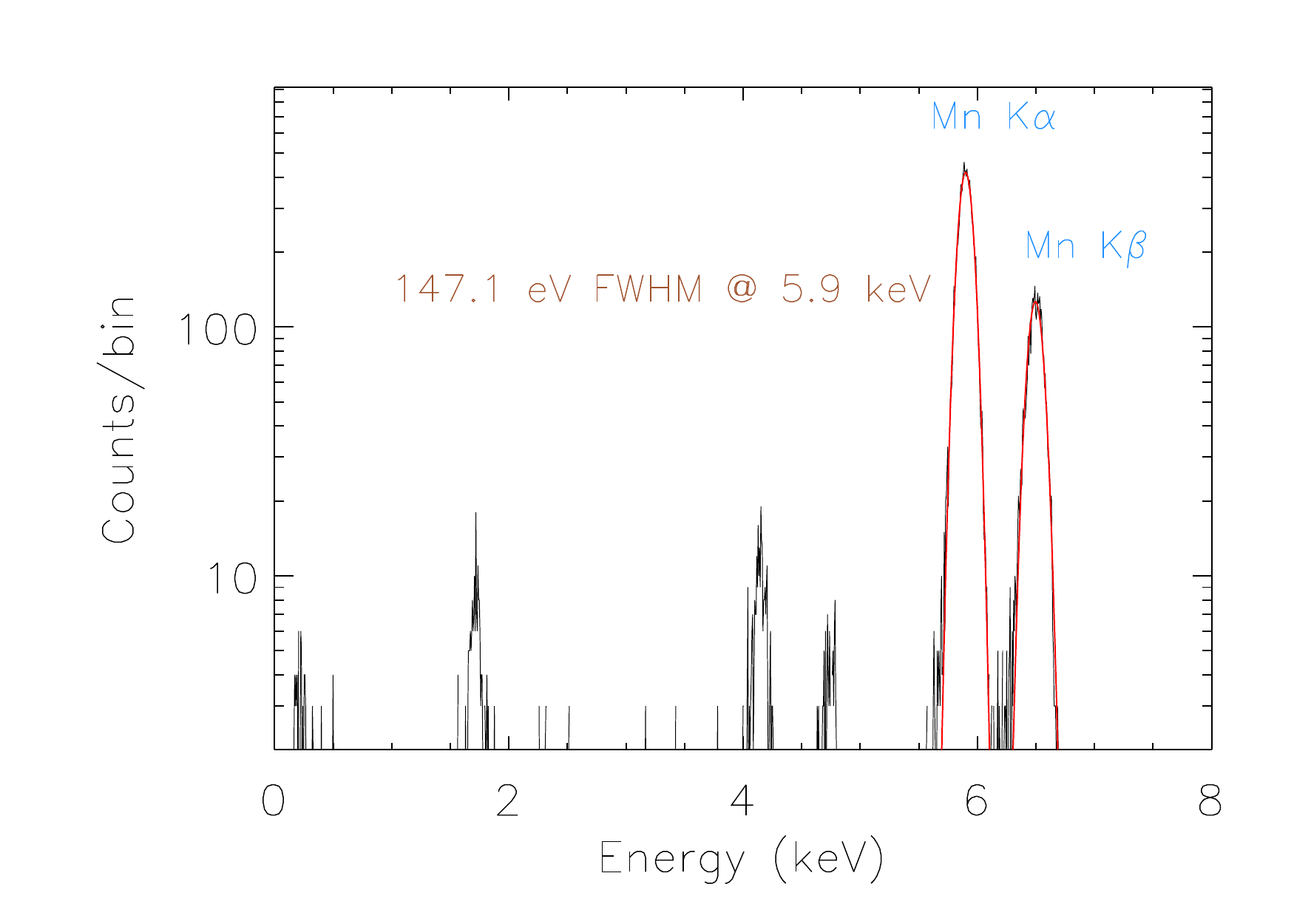}
    \caption{}
    \end{subfigure}
    \caption{(a) Distribution of charge in ADU from the over-scan region. The red dashed line is a Gaussian fit to the
distribution. The read noise is estimated be around 6.44 $e^{-}_{RMS}$ (b) A spectrum showing the Mn K$_\alpha$ (5.9 keV) and K$_\beta$ (6.4 keV) lines from a $^{55}$Fe radioactive source for
single-pixel (grade 0) events. The FWHM is estimated to be 147 eV at 5.9 keV.}
    \label{results}
\end{figure}

Figure \ref{setup}b shows the digital video waveform sampled at every 10 ns as obtained from the Archon controller. The readout speed of the detector is set at 625 kpixel/s (each pixel is 1600 ns long).
The change from the baseline (denoted by the two blue dashed lines) to the signal level (denoted by the two dashed red lines), after the charge packet is transferred from the output gate to the internal channel (see Fig. \ref{sisero}), is directly proportional to the amount of charge transferred.
The signal amplitude is extracted by taking the difference between these two levels (CDS) for each pixel, which are then used to generate the 2D images.   
Read noise is estimated from the distribution of charge in the overclocked pixels (an array of over-scanned pixels with no physical existence). 
An example of such distribution in ADU (Analog to Digital Unit), obtained from one of the dark frames at 250 K (-23$^\circ$C), is shown in Fig. \ref{results}a.
The distribution is fitted with a Gaussian (shown in red dashed line) to quantify the RMS of the distribution. From the known conversion gain of the system, the read noise is estimated to be around 6.44 $e^{-}_{RMS}$.  
An ENC of 6.44 $e^{-}_{RMS}$ in the  buried channel SiSeROs shows a significant improvement in the noise performance compared to the surface channel p-MOSFET SiSeROs, where we estimated the noise to be around 15 $e^{-}_{RMS}$ \citep{chattopadhyay22_sisero}. 

The spectral performance of the device was evaluated using a $^{55}\mathrm{Fe}$ radioisotope. The X-ray images are first corrected for bias (using the over-scan region) and dark current (using the dark frames). We generate an event list of 9-pixel islands around each X-ray event. We then apply a secondary ADU threshold to grade the events and generate spectra for each of the event grades.
An X-ray spectrum showing Mn K$_\alpha$ (5.9 keV) and Mn K$_\beta$ (6.4 keV) lines from the radioactive source is shown in Fig. \ref{results}b. The spectrum is generated using single-pixel events (grade 0 events). The red lines are Gaussian fits to the Mn K$_\alpha$ (5.9 keV) and Mn K$_\beta$ (6.4 keV) lines in the spectra.
We calculate a FWHM of 147 eV at 5.9 keV. As expected, the energy resolution, in the case of buried-channel SiSeROs, has improved significantly from the previous results of 230 eV at 5.9 keV for the surface channel SiSeROs. We also calculated the conversion gain ($G_q$) of the system to be around 800 pA per electron. The transconductance ($g_m$) of the device is estimated to be around 25 $\mu$S.


\section{1/f correlated noise and Digital Filtering}\label{sec:digfiltering}

From the estimated conversion gain of 800 pA per electron and transconductance of 25 $\mu$S, the thermal noise in these SiSeRO devices is expected to be around 2.5-3 $e^{-}_{RMS}$ at a readout speed of 625 kpixel/s. We have observed a large 1/f noise in the frequency spectrum of these detectors with a 1/f corner frequency close to the typical readout rate of the detectors (625 kpixel/s in this case). The Archon uses a default box filter in the CDS of the digital waveform. Although a box filter is preferred in cases where the total noise is dominated by the thermal white noise, it is not ideal when the total noise is dominated by both thermal and 1/f noise. Papers \citep{Stefanov2015} and \citep{Cancelo2012} discuss new digital filtering techniques to reduce the 1/f noise in the sensors. Stefanov et al. 2015 \citep{Stefanov2015} discuss a filter optimized for low frequency correlated noise suppression. On the other hand, Cancelo et al. 2012 \citep{Cancelo2012} estimates the low frequency correlated noise in the waveform and subtract it. Both techniques are shown to provide significant improvement in the overall noise performance. 

In our test setup, we have the flexibility to save the raw waveform for the pixels, which allows us to explore digital filtering techniques. Here, we discuss the effect of an optimized filter (hereafter `cusp') for digital CDS on the noise performance.  
\begin{figure}
    \centering
    \includegraphics[width=0.75\linewidth]{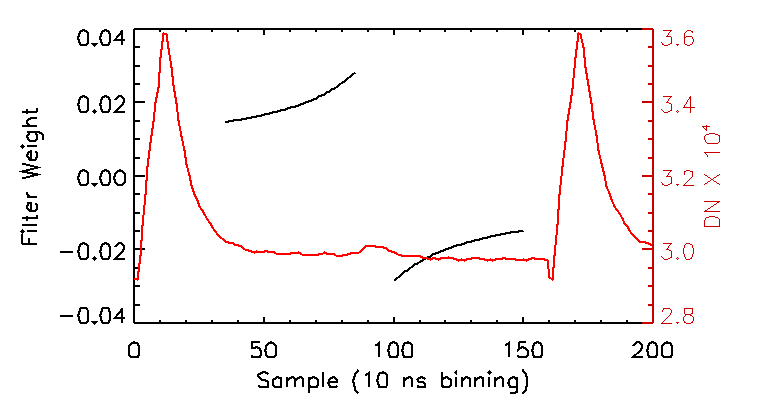}
    \caption{Digital filtering: the black lines (plotted against the left axis) show the weights of the CDS cusp filter used here. The waveform of the SiSeRO is shown in the red line plotted against the right axis. With this filtering technique, the read noise is improved to 6 $e^{-}_{RMS}$.}
    \label{filter}
\end{figure}
In Fig. \ref{filter}, the black solid lines show the weights of the cusp filter in the baseline and signal regions of the waveform (shown in red solid line). The baseline samples multiplied by the positive filter weights and the signal samples multiplied by the negative filter weights are added together to calculate the charge in each pixel. In case of a box CDS, the filter weights are equal along the baseline and signal samples. Because of the correlated low frequency noise embedded in the waveform, the baseline and signals are affected by a small but unequal amount in ADU amplitude, which leads to additional noise after the CDS. To minimize this effect, we apply higher filter gain near the charge transfer, as shown in Fig. \ref{filter}. The noise performance depends on multiple factors including the weights, sampling region etc. While more samples are useful to average out the white thermal noise, the filter peak frequency (highest gain in frequency domain) shifts to a lower frequency (varies roughly as 1/N$_{sample}$) and deeper into the 1/f domain. Therefore, it is important to optimize the sample size and weight ratios simultaneously to achieve best noise performance with this technique. The read noise was improved by 6-7 \% to 6 $e^{-}_{RMS}$ by the cusp filtering from the default box CDS. We expect further improvements in the noise performance by estimating the 1/f correlated noise across a series of pixels, interpolating at the time of the signal for each pixel, and then use this to correct the pixel baseline. This work is currently in progress and will be reported at a later date.

\begin{figure}
    \centering
    \includegraphics[width=0.65\linewidth]{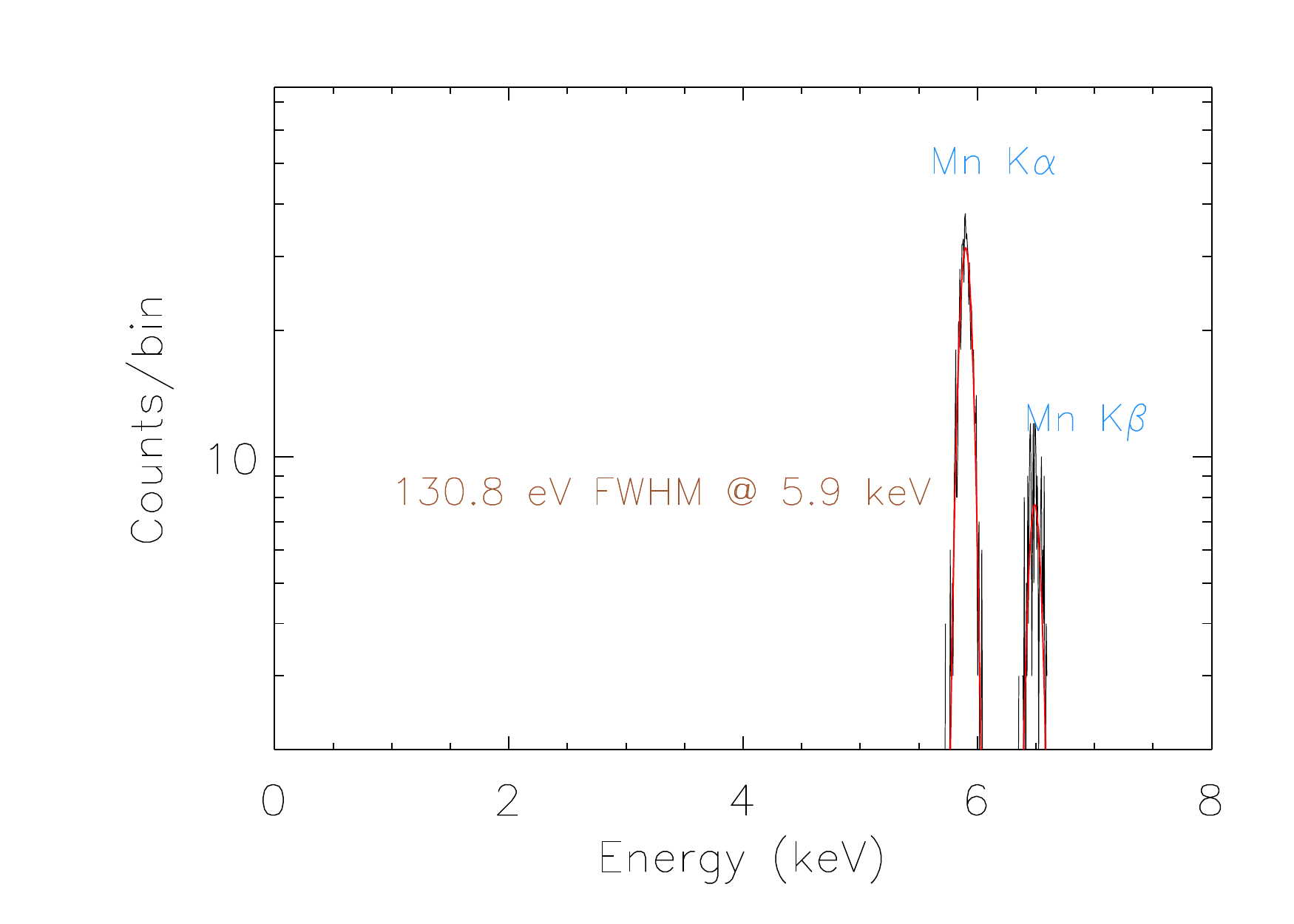}
    \caption{An example spectrum obtained using the digital double baseline CDS filter which improves the read noise to 5.2 $e^{-}_{RMS}$ (see text for more details). It shows the Mn K$_\alpha$ (5.9 keV) and K$_\beta$ (6.4 keV) lines from a $^{55}$Fe radioactive source for
single-pixel (grade 0) events. The FWHM is estimated to be around 130 eV at 5.9 keV.}
    \label{spec_dbl_baseline}
\end{figure}

We also attempted a new digital filtering technique where we include the baseline samples from the adjacent pixel to calculate the baseline for CDS. The advantage of this technique is that, while with more samples from the adjacent pixel the average thermal noise is expected to decrease, this sampling does not also allow the filter peak frequency to shift into the low frequencies, resulting in no excess 1/f noise. The read noise with this technique is estimated to be around 5.2 $e^{-}_{RMS}$. Figure \ref{spec_dbl_baseline} shows a single pixel spectrum obtained from this technique from a $^{55}$Fe radioactive source.     
 The FWHM at 5.9 keV Mn K$_\alpha$ line is estimated to be around 130 eV.


\section{Repetitive non-destructive readout (RNDR)}
\label{sec:rndr}

Since in SieROs, the charge packet remains unaffected in the readout process (not transferred to a doped node like in CCDs), we tried to explore the possibility to utilize RNDR by moving the charge packet multiple
times between the internal channel and the adjacent output gate. This technique has been demonstrated for DEPFET devices with sub-electron read noise yield \citep{wolfel06} and related efforts have been demonstrated on Skipper CCDs \citep{tiffenberg17}. An advantage of repetitive
readout of the same charge signal is that the non-white noise (1/f noise)
can be significantly attenuated, resulting in extremely low noise. 
\begin{figure}
    \centering
     \begin{subfigure}{.62\textwidth}
    \includegraphics[width=\linewidth]{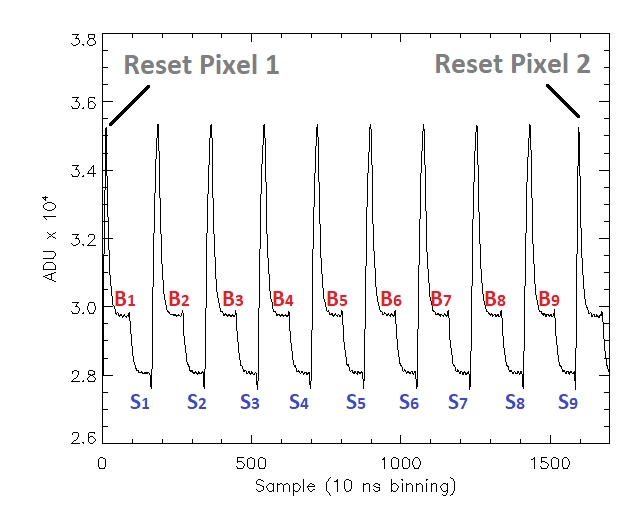}\\
    \caption{}
    \end{subfigure}
    \begin{subfigure}{.474\textwidth}
    \includegraphics[width=\linewidth]{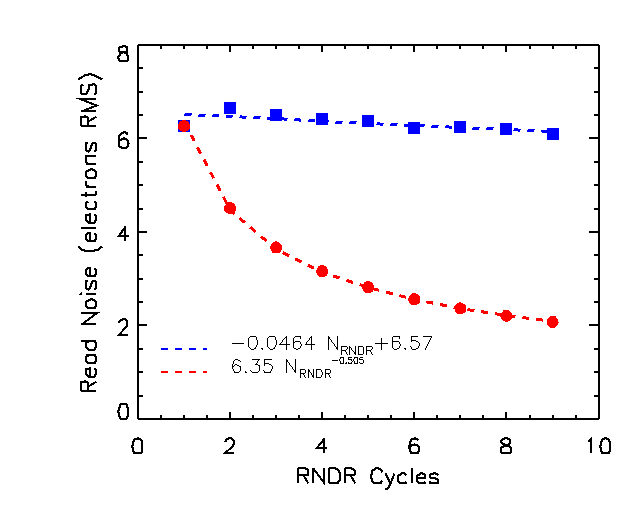}
    \caption{}
    \end{subfigure}
    \begin{subfigure}{.518\textwidth}
    \includegraphics[width=\linewidth]{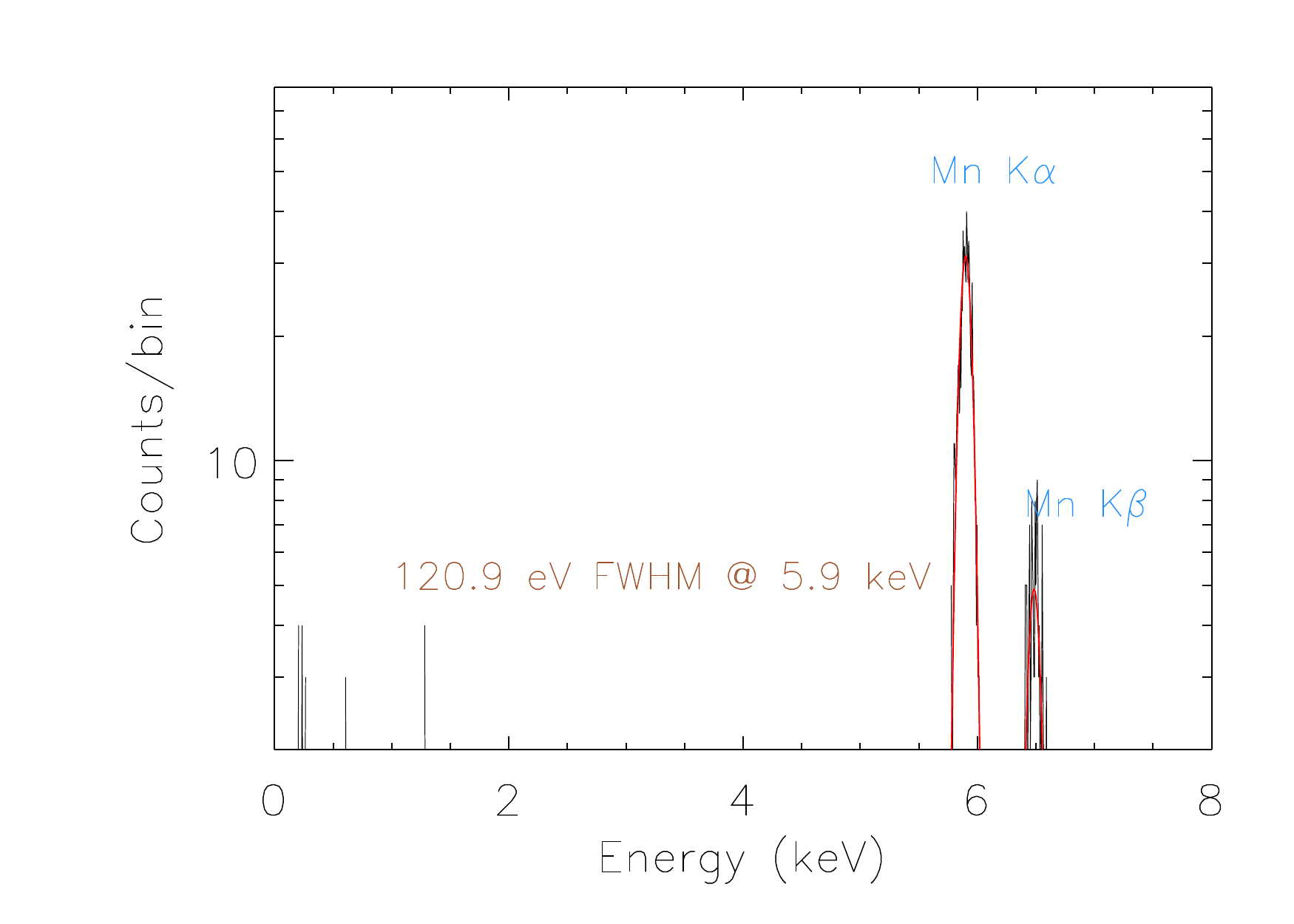}
    \caption{}
    \end{subfigure}
    \caption{Preliminary RNDR results: (a) Waveform for one pixel in RNDR readout process. Here we implemented  nine repetitive readout cycles. (b) Read noise from RNDR analysis. The blue points show the individual read noise measurements from 9 individual repetitive measurements whereas the red data points are obtained when we average out the consecutive measurements. The red data points are shown to follow the expected 1/$\sqrt{N}$ trend. (c) The final spectrum obtained from RNDR analysis showing the Mn K$_\alpha$ (5.9 keV) and K$_\beta$ (6.4 keV) lines from a $^{55}$Fe radioactive source for
single-pixel (grade 0) events. The FWHM is estimated to be around 121 eV at 5.9 keV. The Fano limit for Silicon at 5.9 keV is 119 eV. The read noise for this measurement is 2 $e^{-}_{RMS}$.}
    \label{rndr}
\end{figure}

In Fig. \ref{rndr}a, we show the waveform for one pixel in RNDR readout. Here we implemented nine repetitive readout cycles which results in an increase in the readout time by nine times (overall readout rate is $\sim$63 kpixel/s). For each repetitive cycle,  we first move the charge from the internal channel back to the output gate (OG, see Fig.\ref{sisero}) by applying a positive clock potential ($>$channel potential which is around 2 V) to the OG. Note that in normal applications, OG is connected to a positive DC bias (0.5 V). However, in RNDR, to move the charge back and forth, we clock OG between 0.5 V (OGLow) and 4 V (OGHigh). The last serial clock (Summing Well or SW in CCID93) is then made high and OG is made low simultaneously to move the charge to SW. As OG settles down to OGLow, this introduces the individual baselines (from B$_2$ to B$_9$ in Fig. \ref{rndr}a) that we see in the waveform. The charge packet is then moved back to the internal channel by changing the SW potential to SWLow ($<$OGLow). This introduces the individual signal regions (from S$_2$ to S$_9$ in Fig. \ref{rndr}a) in the waveform.

In Fig. \ref{rndr}b, we show the nine individual (blue squares) and consecutively averaged read noise measurements (red circles). The individual cycles yield similar noise measurements (shown by the blue dashed line fitted to the individual measurements), suggesting that the transfer of the signal charge into and out of the back gate of the transistor is perfect. The red dashed line (${N^\alpha_{RNDR}}$) is a fit to the averaged measurements with a best fit value of -0.505 for $\alpha$. We achieved a read noise as low as 2 $e^{-}_{RMS}$ at the end of the 9$^{th}$ RNDR cycle. 

Figure \ref{rndr}c shows a preliminary spectrum of for Mn K$_\alpha$ (5.9 keV) and K$_\beta$ (6.4 keV) lines generated from the single-pixel events. The FWHM at 5.9 keV is estimated to be around 121 eV. Note that the Fano limit for Silicon at 5.9 keV is 119 eV. 
We plan to refine this technique further such that it helps to reach sub-electron noise levels. The RNDR technique is also relevant to the gain calibration of X-ray detectors  \citep{rodrigues21}.


\section{Summary and future plans}\label{summary}

At SU, we are developing fast low noise readout electronics for the next generation X-ray CCDs for future X-ray astronomy missions. The SiSeRO amplifier, developed by MIT Lincoln Laboratory, is a novel technology for the output stage of X-ray CCDs that can provide very low noise and high speed performance. In this paper, we discussed the working principle of the SiSeROs and characterization test results for a buried-channel SiSeRO device. In our previous paper \cite{chattopadhyay22_sisero}, we used the first generation prototype devices for characterization for which we obtained a read noise of 15 $e^{-}_{RMS}$ and a FWHM energy resolution of 230 eV at 5.9 keV. These earlier prototypes used surface channel transistors, where trapping and de-trapping of charge carriers from the Si interface states degrades the overall noise performance. In the current paper, we characterized a new SiSeRO device with a buried transistor channel. We obtained significant improvements in the noise (around 6 $e^{-}_{RMS}$) and spectral performance (around 150 eV at 5.9 keV). We also demonstrated preliminary results for 1/f correlated noise suppression by using an optimized filter in the digital CDS instead of a default box filter. We have already achieved significant improvement in the noise performance with the use of the new tools: with optimization of the digital filtering, the spectral resolution was improved to 130 eV at 5.9 keV. 

For the first time with such devices, we have explored the RNDR technique. Because the charge packet can be moved back and forth non-destructively between the internal channel and adjacent output gate in the output stage of the SiSeRO, it is possible to apply the RNDR technique to lower the noise significantly, well below the 1/f barrier. In our preliminary analysis, the noise was found to improve from 6 $e^{-}_{RMS}$ to around 2 $e^{-}_{RMS}$ by applying 9 RNDR iterations. The final energy resolution is estimated to be around 121 eV FWHM at 5.9 keV, approaching the Fano limit of Silicon at these energies.   

The results shown here are extremely encouraging. In future work, we plan to mature the digital filtering tools and develop techniques to accurately estimate the 1/f correlated noise and subtract it from the waveform. These new techniques are expected to improve the noise performance of the devices even further. With the 1/f noise subtraction and further improvements in RNDR, it might be possible to achieve sub-electron noise performance with even first generation SiSeRO devices. One of the goals of our ongoing SiSeRO development program is to characterize larger arrays of SiSeROs with multiple parallel readouts; the first examples will be fabricated in the near future. We have developed an ASIC-based readout system at SU to enable parallel readout of these devices, the details of which can be found in \citep{herrmann20_mcrc} and other papers submitted in this conference \citep{Oreletal2022}. Apart from the improvements in the readout electronics and analysis software, we plan to continue testing the buried-channel devices in combination with detailed device simulations to mature the SiSeRO technology.  

\acknowledgments 

This work has been supported by NASA grants APRA 80NSSC19K0499 ``Development
of Integrated Readout Electronics for Next Generation X-ray CCDs” and SAT
80NSSC20K0401 ``Toward Fast, Low-Noise, Radiation-Tolerant X-ray Imaging Arrays for
Lynx: Raising Technology Readiness Further.”


\begin{thebibliography}{10}

\bibitem{Lesser15_ccd}
Lesser, M., ``A summary of charge-coupled devices for astronomy,'' {\em
  Publications of the Astronomical Society of the Pacific}~{\bf 127}(957),
  1097 (2015).

\bibitem{gruner02_ccd}
Gruner, S.~M., Tate, M.~W., and Eikenberry, E.~F., ``Charge-coupled device area
  x-ray detectors,'' {\em Review of Scientific Instruments}~{\bf 73}(8),
  2815--2842 (2002).

\bibitem{gaskin15_lynx}
{Gaskin}, J.~A., {Weisskopf}, M.~C., {Vikhlinin}, A., {Tananbaum}, H.~D.,
  {Bandler}, S.~R., {Bautz}, M.~W., {Burrows}, D.~N., {Falcone}, A.~D.,
  {Harrison}, F.~A., {Heilmann}, R.~K., {Heinz}, S., {Hopkins}, R.~C.,
  {Kilbourne}, C.~A., {Kouveliotou}, C., {Kraft}, R.~P., {Kravtsov}, A.~V.,
  {McEntaffer}, R.~L., {Natarajan}, P., {O'Dell}, S.~L., {Petre}, R.,
  {Prieskorn}, Z.~R., {Ptak}, A.~F., {Ramsey}, B.~D., {Reid}, P.~B., {Schnell},
  A.~R., {Schwartz}, D.~A., and {Townsley}, L.~K., ``{The X-ray Surveyor
  Mission: a concept study},'' in [{\em UV, X-Ray, and Gamma-Ray Space
  Instrumentation for Astronomy XIX}{\nolinebreak\hspace{0.1em}]},  {\em
  Society of Photo-Optical Instrumentation Engineers (SPIE) Conference Series}
  {\bf 9601},  96010J (Aug. 2015).

\bibitem{bautz18}
Bautz, M., Foster, R., LaMarr, B., Malonis, A., Prigozhin, G., Miller, E.,
  Grant, C.~E., Burke, B., Cooper, M., Craig, D., Leitz, C., Schuette, D., and
  Suntharalingam, V., ``{Toward fast low-noise low-power digital CCDs for Lynx
  and other high-energy astrophysics missions},'' in [{\em Space Telescopes and
  Instrumentation 2018: Ultraviolet to Gamma Ray}{\nolinebreak\hspace{0.1em}]},
   den Herder, J.-W.~A., Nikzad, S., and Nakazawa, K., eds.,  {\bf 10699},  238
  -- 248, International Society for Optics and Photonics, SPIE (2018).

\bibitem{bautz19}
{Bautz}, M.~W., {Burke}, B.~E., {Cooper}, M., {Craig}, D., {Foster}, R.~F.,
  {Grant}, C.~E., {LaMarr}, B.~J., {Leitz}, C., {Malonis}, A., {Miller}, E.~D.,
  {Prigozhin}, G., {Schuette}, D., {Suntharalingam}, V., and {Thayer}, C.,
  ``{Toward fast, low-noise charge-coupled devices for Lynx},'' {\em Journal of
  Astronomical Telescopes, Instruments, and Systems}~{\bf 5},  021015 (Apr.
  2019).

\bibitem{bautz20}
Bautz, M., Burke, B., Cooper, M., Craig, D., Donlon, K., Foster, R., Grant,
  C.~E., LaMarr, B., Leitz, C., Malonis, A., Miller, E., Prigozhin, G., Thayer,
  C., Allen, S., Herrmann, S., Chattopadhyay, T., and Morris, R.~G.,
  ``{Progress toward fast, low-noise, low-power CCDs for Lynx and other
  high-energy astrophysics missions},'' in [{\em Space Telescopes and
  Instrumentation 2020: Ultraviolet to Gamma Ray}{\nolinebreak\hspace{0.1em}]},
   den Herder, J.-W.~A., Nikzad, S., and Nakazawa, K., eds.,  {\bf 11444},
  1318 -- 1323, International Society for Optics and Photonics, SPIE (2020).

\bibitem{chattopadhyay22_ccd}
Chattopadhyay, T., Herrmann, S.~C., Orel, P., Morris, R.~G., Prigozhin, G.~Y.,
  Malonis, A.~C., Foster, R.~F., Craig, D.~M., Burke, B.~E., Allen, S.~W., and
  Bautz, M.~W., ``{Development and characterization of a fast and low noise
  readout for the next generation x-ray charge-coupled devices},'' {\em Journal
  of Astronomical Telescopes, Instruments, and Systems}~{\bf 8}(2),  1 -- 12
  (2022).

\bibitem{herrmann20_mcrc}
Herrmann, S., Wong, J., Chattopadhyay, T., Morris, R.~G., Burke, B., Prigozhin,
  G., Cooper, M., Craig, D., Donlon, K., Foster, R., Malonis, A., Bautz, M.,
  and Allen, S., ``{MCRC V1: development of integrated readout electronics for
  next generation x-ray CCD detectors for future satellite observatories},'' in
  [{\em X-Ray, Optical, and Infrared Detectors for Astronomy
  IX}{\nolinebreak\hspace{0.1em}]},  Holland, A.~D. and Beletic, J., eds.,
  {\bf 11454},  412 -- 418, International Society for Optics and Photonics,
  SPIE (2020).

\bibitem{Bautzetal2022}
{Bautz}, M.~W., {Foster}, R., {Grant}, C.~E., {LaMarr}, B., {Malonis}, A.,
  {Miller}, E.~D., {Prigozhin}, G., {Burke}, B., {Cooper}, M., {Donlon}, K.,
  {Lambert}, R., {Warner}, K., {Young}, D., {Chattopadhyay}, T., {Herrmann},
  S., {Morris}, R.~G., {Leitz}, C., and {Allen}, S., ``{Performance of high
  frame-rate CCDs for future strategic missions},'' in [{\em Space Telescopes
  and Instrumentation 2022: Ultraviolet to Gamma
  Ray}{\nolinebreak\hspace{0.1em}]},  {\em Society of Photo-Optical
  Instrumentation Engineers (SPIE) Conference Series} {\bf 12181},  12181--85
  (2022).

\bibitem{Oreletal2022}
{Orel}, P., {Hermann}, S.~C., {Chattopadhyay}, T., {Morris}, R.~G.,
  {Prigozhin}, G., {Cooper}, M.~J., {Donlon}, K., {Foster}, R., {Malonis}, A.,
  {Allen}, S.~W., and {Bautz}, M.~W., ``{X-ray speed reading with the MCRC: a
  low noise CCD readout ASIC enabling readout speeds of 5 Mpixel/s/channel},''
  in [{\em X-Ray, Optical, and Infrared Detectors for Astronomy
  X}{\nolinebreak\hspace{0.1em}]},  {\em Society of Photo-Optical
  Instrumentation Engineers (SPIE) Conference Series} {\bf 12191},  12191--73
  (2022).

\bibitem{chattopadhyay22_sisero}
Chattopadhyay, T., Herrmann, S., Burke, B.~E., Donlon, K., Prigozhin, G.,
  Morris, G., Orel, P., Cooper, M., Malonis, A., Wilkins, D.~R.,
  Suntharalingam, V., Allen, S.~W., Bautz, M.~W., and Leitz, C., ``{First
  results on SiSeRO devices: a new x-ray detector for scientific
  instrumentation},'' {\em Journal of Astronomical Telescopes, Instruments, and
  Systems}~{\bf 8}(2),  1 -- 12 (2022).

\bibitem{kemmer87_depfet}
Kemmer, J. and Lutz, G., ``New detector concepts,'' {\em Nuclear Instruments
  and Methods in Physics Research Section A: Accelerators, Spectrometers,
  Detectors and Associated Equipment}~{\bf 253}(3),  365--377 (1987).

\bibitem{strueder00_depfet_imager}
Strueder, L., Meidinger, N., Pfeffermann, E., Hartmann, R., Braeuninger, H.~W.,
  Krause, N., Hartner, G.~D., Dennerl, K., Haberl, F., Kemmer, S., Popp, M.,
  Truemper, J.~E., Kollmer, J., Johannes, T., Lutz, G., Hauff, D., Richter,
  R.~H., Klein, P., Hoernel, N., Solc, P., Eckhardt, R., Fischer, P., Neeser,
  W., Ulrici, J., Wermes, N., Holl, P., Lechner, P., Kemmer, J., Soltau, H.,
  Stoetter, R., Weber, U., and Weichert, U., ``{Fully depleted
  backside-illuminated spectroscopic active pixel sensors from the infrared to
  x rays (1 eV to 25 keV)},'' in [{\em X-Ray Optics, Instruments, and Missions
  III}{\nolinebreak\hspace{0.1em}]},  Truemper, J.~E. and Aschenbach, B., eds.,
   {\bf 4012},  200 -- 217, International Society for Optics and Photonics,
  SPIE (2000).

\bibitem{matsunaga91}
Matsunaga, Y., Yamashita, H., and Ohsawa, S., ``A highly sensitive on-chip
  charge detector for ccd area image sensor,'' {\em IEEE Journal of Solid-State
  Circuits}~{\bf 26}(4),  652--656 (1991).

\bibitem{chattopadhyay20_spie}
Chattopadhyay, T., Herrmann, S., Allen, S., Hirschman, J., Morris, G., Bautz,
  M., Malonis, A., Foster, R., Prigozhin, G., Craig, D., and Burke, B.,
  ``{Tiny-box: a tool for the versatile development and characterization of low
  noise fast x-ray imaging detectors},'' in [{\em X-Ray, Optical, and Infrared
  Detectors for Astronomy IX}{\nolinebreak\hspace{0.1em}]},  Holland, A.~D. and
  Beletic, J., eds.,  {\bf 11454},  368 -- 385, International Society for
  Optics and Photonics, SPIE (2020).

\bibitem{archon14}
Bredthauer, G., ``{Archon: A modern controller for high performance
  astronomical CCDs},'' in [{\em Ground-based and Airborne Instrumentation for
  Astronomy V}{\nolinebreak\hspace{0.1em}]},  Ramsay, S.~K., McLean, I.~S., and
  Takami, H., eds.,  {\bf 9147},  1730 -- 1740, International Society for
  Optics and Photonics, SPIE (2014).

\bibitem{Stefanov2015}
Stefanov, K., ``Digital {CDS} for image sensors with dominant white and 1/f
  noise,'' {\em Journal of Instrumentation}~{\bf 10},  P04003--P04003 (apr
  2015).

\bibitem{Cancelo2012}
{Cancelo}, G.~I., {Estrada}, J.~C., {Moroni}, G.~F., {Treptow}, K., {Zmuda},
  T., and {Diehl}, H.~T., ``Achieving sub electron noise in ccd systems by
  means of digital filtering techniques that lower 1/f pixel correlated
  noise,'' {\em Exp Astron}~{\bf 34},  13--29 (apr 2012).

\bibitem{wolfel06}
Wölfel, S., Herrmann, S., Lechner, P., Lutz, G., Porro, M., Richter, R.,
  Strüder, L., and Treis, J., ``Sub-electron noise measurements on repetitive
  non-destructive readout devices,'' {\em Nuclear Instruments and Methods in
  Physics Research Section A: Accelerators, Spectrometers, Detectors and
  Associated Equipment}~{\bf 566}(2),  536--539 (2006).

\bibitem{tiffenberg17}
Tiffenberg, J., Sofo-Haro, M., Drlica-Wagner, A., Essig, R., Guardincerri, Y.,
  Holland, S., Volansky, T., and Yu, T.-T., ``Single-electron and single-photon
  sensitivity with a silicon skipper ccd,'' {\em Phys. Rev. Lett.}~{\bf 119},
  131802 (Sep 2017).

\bibitem{rodrigues21}
{Rodrigues}, D., {Andersson}, K., {Cababie}, M., {Donadon}, A., {Botti}, A.,
  {Cancelo}, G., {Estrada}, J., {Fernandez-Moroni}, G., {Piegaia}, R.,
  {Senger}, M., {Haro}, M.~S., {Stefanazzi}, L., {Tiffenberg}, J., and
  {Uemura}, S., ``{Absolute measurement of the Fano factor using a
  Skipper-CCD},'' {\em Nuclear Instruments and Methods in Physics Research
  A}~{\bf 1010},  165511 (Sept. 2021).

\end{thebibliography}

\end{document}